\documentclass[prb,twocolumn,showpacs]{revtex4}

\usepackage{epsfig}

\newcommand {\be} {\begin{equation}}   
\newcommand {\ee} {\end{equation}}
\newcommand {\bea} {\begin{eqnarray}}
\newcommand {\eea} {\end{eqnarray}}
\newcommand {\bes} {\begin{displaymath}}
\newcommand {\ees} {\end{displaymath}}
\newcommand {\beas} {\begin{eqnarray*}}
\newcommand {\eeas} {\end{eqnarray*}}
\newcommand {\w} {\omega}
\newcommand {\ecpp} {\varepsilon_{{\rm c}\parallel}}
\newcommand {\ecpr} {\varepsilon_{{\rm c}\perp}}
\newcommand {\emed} {\varepsilon_{{\rm m}}}
\newcommand {\ein} {\varepsilon_{{\rm in}}}
\newcommand {\epp} {\varepsilon_{\parallel}}

\newcommand {\ei} {\varepsilon_i}

\begin{document}

\title{Resonant coupling between localized plasmons and
anisotropic molecular coatings in ellipsoidal metal nanoparticles}
\author{Tobias Ambj\"ornsson}
\affiliation{NORDITA (Nordic Institute for
Theoretical Physics), Blegdamsvej 17, DK-2100 Copenhagen \O, Denmark.}
\email[Corresponding author, e-mail: ]{ambjorn@nordita.dk}
\author{Gautam Mukhopadhyay}
\affiliation{Physics Department, Indian
Institute of Technology, Bombay, Powai, Mumbai 400076, India.}
\author{S. Peter Apell}
\affiliation{Department of Applied Physics,
 Chalmers University of Technology \\ and G\"oteborg University, 412
 96 G\"oteborg, Sweden.}
 \author{Mikael K\"all }
\affiliation{Department of Applied Physics,
 Chalmers University of Technology \\ and G\"oteborg University, 412
 96 G\"oteborg, Sweden.}

\begin{abstract}
We present an analytic theory for the optical properties of
ellipsoidal plasmonic particles covered by anisotropic molecular
layers. The theory is applied to the case of a prolate spheroid
covered by chromophores oriented parallel and perpendicular to the
metal surface. For the case that the molecular layer resonance
frequency is close to being degenerate with one of the particle
plasmon resonances strong hybridization between the two resonances
occur. Approximate analytic expressions for the hybridized resonance
frequencies, their extinction cross section peak heights and widths
are derived. The strength of the molecular - plasmon interaction is
found to be strongly dependent on molecular orientation and suggest
that this sensitivity could be the basis for novel nanoparticle based
bio/chemo-sensing applications.
\end{abstract}


\maketitle

\section{Introduction} 

The recent decade has witnessed extensive research
efforts directed at localized surface plasmons (LSP's), i.e resonant
charge-density oscillations confined to sub-wavelength metal
structures, such as particles\cite{Kreibig}, holes\cite{Prikulis04},
shells\cite{HalasScience} or rods\cite{ElSayed1999}.  LSP's are
important because they lead to strongly enhanced optical absorption
and scattering cross-sections, and because they readily couple to
optical far-fields, unlike the ordinary surface plasmons of extended
metal surfaces.  In the small particle regime, the LSP resonance
wavelengths of a single nanostructure are uniquely determined by its
shape and dielectric function, and by the optical constants of the
embedding medium.  Thus, the color of the nanostructure can be tuned
over an extended wavelength-range, including most of the visible and
infrared regions in the case of silver or gold structures, through a
variation in shape. Moreover, the color can be used to "sense" the
optical properties of the surrounding. In addition, excitation of
LSP's leads to strongly enhanced and localized fields in certain
regions near the metal surface, and these fields can be used to
amplify various molecular cross-sections.  A bi-sphere system, for
example, support resonances for which the field is concentrated to the
gap between the spheres, which is crucial for single-molecule
surface-enhanced Raman scattering (SERS) \cite{XuPRB2000}, and a sharp
point or protrusion support LSP-enhanced fields at its apex, which can
be utilized for various types of near-field microscopy
\cite{BouhelierPRL2003}.  Sensing applications rely on the fact that
it is only the optical properties of the material within the zone of
high field-enhancement that strongly affect the LSP resonance
condition.\cite{HaesJPCB2004}  The LSP spectrum, which can be
measured through far-field extinction or Rayleigh scattering
spectroscopy, can therefore be used for ultrasensitive sensing
applications aimed, for example, at quantifying various biomolecular
recognition reactions.\cite{HaesNL2004, DahlinJACS2005}

As is known from a range of experiments and calculations on
nanostructured or flat metal surfaces covered with chromophores, such
as dye molecules, the interaction between a plasmonic structure and
its dielectric environment can be very strong if the dielectric
possesses an optical resonance that is degenerate with a surface
plasmon resonance mode.\cite{Pockrand1978, Glass1980, Craighead1981,
Wang1982, Kotler1982, Bellessa2004,
Wiederrecht_Wurtz_etal,Dintinger2005} This strong coupling also forms
the basis for important applications in molecular spectroscopy, in
particular surface-enhanced resonance Raman scattering (SERRS) and
surface-enhanced fluorescence.\cite{Moskovits1985, Itoh2003, Xu2004}
For atoms and molecules confined in microcavities similar strong
coupling can occur.\cite{Agarwal,Armitage_Skolnick} In many
applications, it is advantageous to bind the molecular layer to the
metal via specific functional groups, for example thiols, in which
case the molecular transition dipole moment will have a more or less
well-defined orientation relative to the local fields generated at
plasmon resonance. This will in turn affect the degree of coupling
between a molecular resonance and a plasmon.

In this work, we present an analytical theory for calculating the
optical properties of a sub-wavelength metallic ellipsoid, the
prototypical example of a nanostructure supporting tunable LSP's,
covered by an optically anisotropic molecular layer. We then utilize
this formalism for investigating strong coupling effects in
nano-plasmonics. The quasi-static theory of dipolar plasmons for
ellipsoids has played an important role in nano-optics, as it can be
used to model a wide range of nanoparticle shapes of practical
interest, including rods, spheres, oblate and prolate particles,
accessing a broad range of frequencies. The extension of this theory
to ellipsoidal cores with anisotropic coatings means that it is now possible
to also investigate the effect of molecular orientation relative to
the core surface.

\section{Theory} 

\begin{figure}
  \scalebox{0.42}{\includegraphics{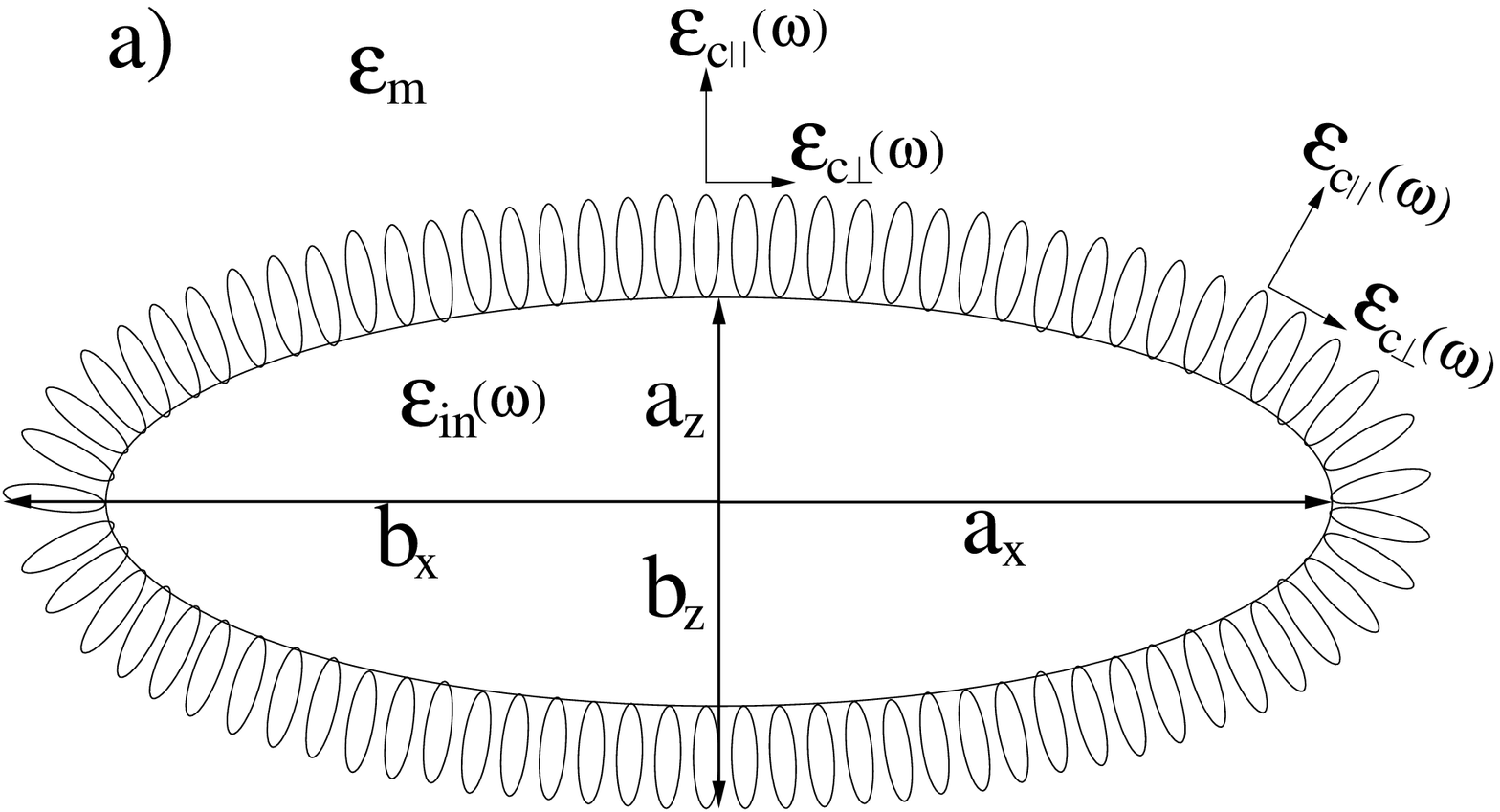}}
  \scalebox{0.41}{\includegraphics{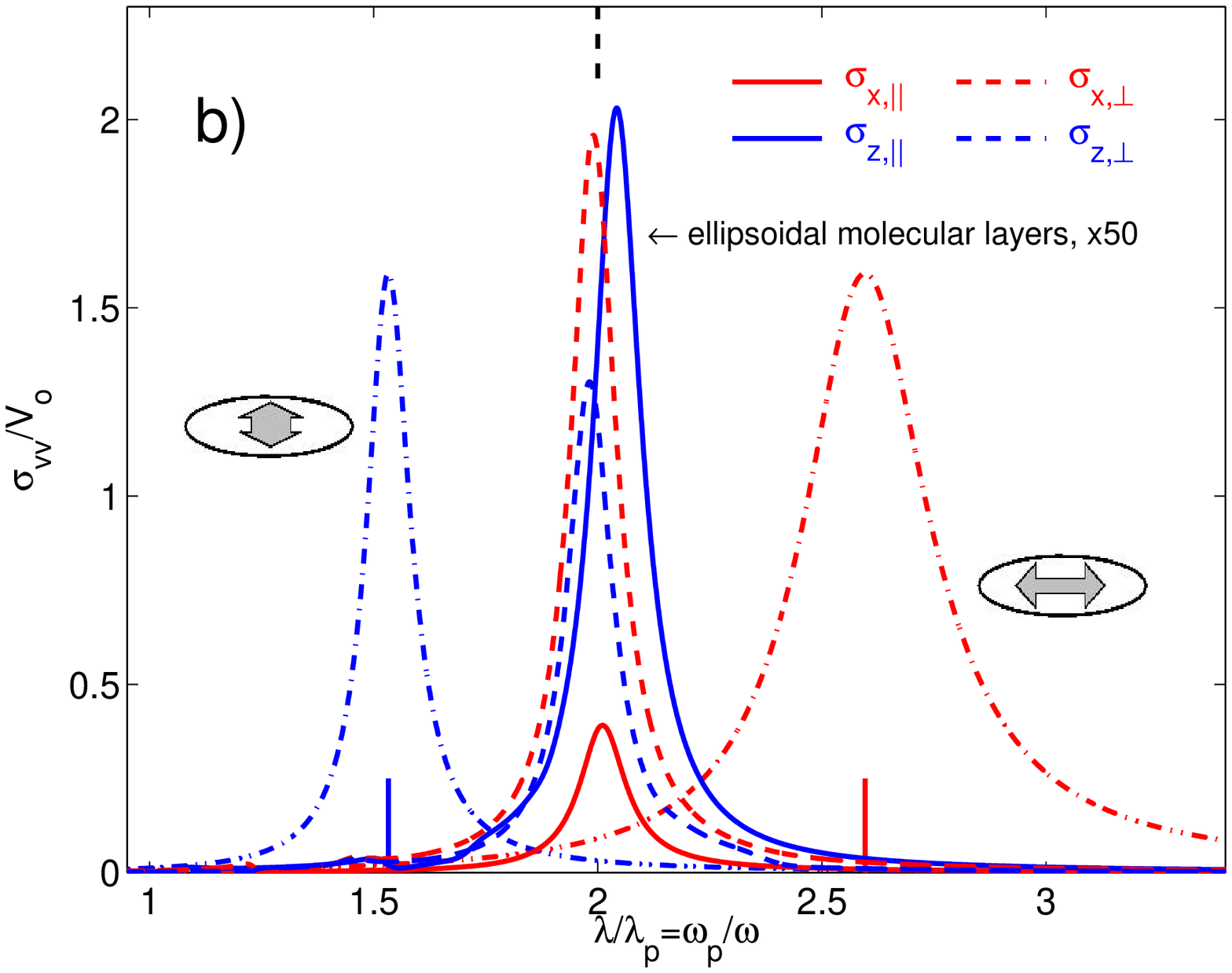}}
  \caption{  
    a) A cut through a coated ellipsoidal
    nanoparticle. The principal semiaxes perpendicular to the paper
    are $a_y$ and $b_y$ respectively. The relevant dielectric
    functions $\epsilon_{in}(\w)$, $\varepsilon_{c\|}(\w)$,
    $\varepsilon_{c\perp}(\w)$ and $\varepsilon_m$ are also indicated.
    b) Extinction spectrum, where $\sigma_{vv}=\w \Im [\alpha_{vv}]$,
    $v=x,y,z$. The dash-dotted curves are for an {\em uncoated}
    metallic ellipsoid, where the red (blue) curve is for the external
    field along the $x$-axis ($z$-axis). The red solid ($x$-direction)
    and blue solid ($z$-direction) curves correspond to the case of
    an ellipsoidal molecular layer ({\em without} an interior metallic
    region, i.e. we set $\epsilon_{i}=1$) with a resonance {\em
    parallel} to the layer normal. Here, this situation is modeled by
    taking $\gamma_{0\|}/v_0=0.2$, $\gamma_{\infty\|}/v_0=0.02$, and
    $\gamma_{0\perp}/v_0=\gamma_{\infty\perp} /v_0=0.01$
    (corresponding to effective oscillator strength $c_{0\|}\approx
    2.26$, and large frequency refractive index ${\rm n}_\|\approx
    1.06$, see text). The red dashed ($x$-direction) and blue dashed
    ($z$-direction) curves are for the case that the molecular layer
    has a resonance {\em perpendicular} to the layer normal. We take
    $\gamma_{0\perp}/v_0=0.1$ and
    $\gamma_{\infty\perp}/v_0=\gamma_{0\|}/v_0=\gamma_{\infty\|}
    /v_0=0.01$ (so that $c_{0\perp}\approx 1.13$ and ${\rm
    n}_\perp\approx 1.07$). The following parameters were used:
    $e_o^2=0.8$, $s=1$, $\Delta_V=0.08$, $\Gamma_p/\w_p=0.05$,
    $\w_{0\|}/\w_p=\w_{0\perp} /\w_p=0.5$,
    $\Gamma_\|/\w_p=\Gamma_\perp/\w_p=0.03$ and $\varepsilon_m=1$. The
    extinction cross-sections are shown in units of the total volume
    $V_o$. The solid and dashed vertical bars indicate the plasmon and
    molecular resonance frequencies, respectively. Notice the small
    peak heights of the molecular layer resonances (here enlarged by a
    factor 50) compared with the plasmon resonance heights.}
  \label{fig:particle}
\end{figure}

While the polarizability of a metallic particle with
and without coating has been studied in the past, and can be found in
text books, to our knowledge, the effect of anisotropy, \textit{i.e.},
molecular orientation, has not been analyzed earlier in the context of
plasmonics.  Here, we recapitulate recent analytic
results\cite{Ambjornsson_Mukhopadhyay,Ambjornsson_Apell_Mukhopadhyay}
for the dipolar polarizability of an ellipsoid with an anisotropic
coating (the coating dielectric function being different parallel and
perpendicular to the coating normal, see Fig. \ref{fig:particle}a),
and combine these results with realistic microscopic dielectric
functions for the metallic nanoparticle and the coating.

The system we have in mind is depicted in Fig. \ref{fig:particle}a: We
consider a coated ellipsoidal particle in an external electric field
$\vec{E}_0$, with field-component $E_{0v}$ in the $v$-direction
($v=x,y,z$). The frequency of the external field is $\w=2\pi
c/\lambda$, where $c$ is the speed of light and $\lambda$ is the
wavelength. The principal semiaxes of the inner and outer ellipsoids
are $a_v$ and $b_v$ with $a_x\ge a_y\ge a_z$ and $b_x\ge b_y\ge
b_z$. The shape of the coated particle is completely specified by the
ellipticity of the inner surface $e_i^2\equiv 1-a_z^2/a_x^2$, the
ellipticity of the outer surface $e_o^2\equiv 1-b_z^2/b_x^2$
($e_i^2=e_o^2=0$ for a sphere) and $s\equiv
(a_x^2-a_y^2)/(a_x^2-a_z^2)$; $e_i^2$, $e_o^2$ and $s$ are all in the
range {[}0,1{]}. The coating thickness is determined by the relative
coating volume $\Delta_V=V_{\rm coat}/V_o$, where $V_o=4\pi b_x b_y
b_z/3$ is the total ellipsoidal volume and $V_{\rm coat}=V_o-V_i$ is
the coating volume ($V_i=4\pi a_x a_y a_z/3$ is the volume of the
inner ellipsoid). For the case of a thin coating the ellipticities are
related: $e_i^2=e_o^2(1+\bar{\delta})$, where $\bar{\delta}$ is the
relative coating thickness parameter (see Refs.
\onlinecite{Ambjornsson_Mukhopadhyay,Ambjornsson_Apell_Mukhopadhyay});
the relative coating volume (a dimensionless entity) introduced above
can be written in terms of the parameter $\bar{\delta}$ according to
$\Delta_V=F_o\bar{\delta}/2$, where
$F_o=1+(1-se_o^2)^{-1}+(1-e_o^2)^{-1}$ (see Ref.
\onlinecite{Ambjornsson_Apell_Mukhopadhyay}).  The ``material''
properties of the coated nanoparticle enters through the relevant
dielectric functions: We denote the dielectric function of the inner
ellipsoid by $\ein(\w)$. The coating has dielectric function
$\ecpp(\w)$ in the normal direction and $\ecpr(\w)$ in the tangential
direction. The dielectric function of the surrounding medium is
assumed to be real and frequency independent and is denoted by
$\emed$.  The entities above completely determine the electromagnetic
response of a coated ellipsoidal particle, i.e., determine, for
instance, the particle polarizability $\alpha_{vv}$ and the electric
field distribution in and around the coated
ellipsoid.\cite{Ambjornsson_Mukhopadhyay,Ambjornsson_Apell_Mukhopadhyay}

The full expression for the dipolar polarizability $\alpha_{vv}$
(defined through $p_v=4\pi\varepsilon_0\varepsilon_m \alpha_{vv}E_{0v}$,
where $p_v$ is the $v$-component, $v=x,y$ or $z$, of the induced
dipole moment and $\varepsilon_0$ is the permittivity of free space)
in terms of the geometric and dielectric entities above is given in
Refs. \onlinecite{Ambjornsson_Mukhopadhyay} and
\onlinecite{Ambjornsson_Apell_Mukhopadhyay}: A knowledge of $\alpha_{vv}$
requires a function $H_v(u,s;q)$, satisfying Heun's equation
\cite{Ronveaux,Slavyanov_Lay},
evaluated at the inner, $q=e_i^2$, and the outer, $q=e_o^2$,
surfaces. The shape parameter $s$ was introduced above. The anisotropy in
the dielectric function of the coating ($\varepsilon_{c\|}$ and
$\varepsilon_{c\perp}$) enters through $u$ defined as
 \be
u=u_\pm=-\frac{1}{2}\pm \frac{1}{2}(1+8\varepsilon_{\perp}
\varepsilon_{\parallel}^{-1})^{1/2},\label{eq:u_main}
  \ee
where $\varepsilon_{\mu}(\w)=\varepsilon_{c\mu}(\w)/\varepsilon_m$
($\mu=\perp$ or $\|$).  For the case of spheroids (two of the
principal axes are equal) $H_v(u,s; q)$ is expressible in terms of
hypergeometric functions, which are available in standard mathematical
numerical packages, such as MatLab or Mathematica. For general
ellipsoids $H_v(u,s; q)$ is straightforwardly generated using a
recurrence relation, explicitly given in
Ref. \onlinecite{Ambjornsson_Mukhopadhyay}. Introducing a second
function $r_v(u,s; q)$ ($v$=$x$, $y$ or $z$), required while implementing
the boundary conditions and defined in terms of
$H_v(u,s;q)$ according to
  \be
r_v (u,s; q)= 1-f_v (s;q)\{ 1-u+2q\frac{\partial}{\partial q}
\ln {[} H_v (u,s; q){]}\},\label{eq:r_v}
  \ee
where $f_x(s;q) =1$, $f_y (s;q)=1-sq$ and $f_z (s;q)=1-q$, the
polarizability of an ellipsoid with an anisotropic coating is
  \be   
\alpha_{vv}=\frac{V_o}{4\pi
n_v^o}\frac{I_v(\kappa=-1)}{I_v(\kappa=1/n_v^o-1)}\label{eq:alpha1},
  \ee
where
  \bea
I_v(\kappa )&=&{[}r^o_v(u_+)+\kappa
\epp^{-1}{]}{[}r^i_v(u_-)-\ei\epp^{-1}{]}\nonumber\\
& &-\rho_v {[}
r^o_v(u_- )+\kappa \epp^{-1}{]}{[}r^i_v(u_+ )-\ei \epp^{-1}{]}.\label{eq:alpha2}
  \eea
where $\varepsilon_i=\varepsilon_{\rm in}/\varepsilon_m$, and we have
above introduced the short-hand notation $r_v^i(u_\pm )\equiv
r_v(u_\pm ,s;e_i^2)$ and $r_v^o(u_\pm )\equiv r_v(u_\pm ,s;
e_o^2)$. $n^o_v$ is the standard depolarization factor for the
$v$-direction for the outer surface and satisfies the sum 
rule\cite{Landau_Lifshitz_ecm,Ambjornsson_Mukhopadhyay,Ambjornsson_Apell_Mukhopadhyay}:
$n_x^o+n_y^o+n_z^o=1$.
We have also introduced
  \be
\rho_v\equiv (\frac{e_o}{e_i})^{(u_+-u_-)}\frac{H_v^o(u_-)H_v^i(u_+)}
{H_v^o(u_+)H_v^i(u_-)},\label{eq:rho}
  \ee 
where $H_v^o(u_\pm)\equiv H_v(u_\pm,s;e_o^2)$ and $H_v^i(u_\pm)\equiv
H_v(u_\pm,s;e_i^2)$.  We note that the total volume $V_o$ only enters
as a prefactor in the expression for $\alpha_{vv}$.  The geometry of
the particle enters through the geometric entities $n_v^o$,
$r_v^i(u_\pm)$ and $r_v^o (u_\pm )$. The standard isotropic
depolarization factor $n_v^o$ depends only on the shape ($s$ and $e_o
^2$), whereas $r_v^i(u_\pm )$ and $r_v^o (u_\pm)$ couple the geometry
to the dielectric asymmetry of the coating. For an anisotropic coated
sphere the expression for the polarizability agrees with the result
obtained in Ref. \onlinecite{Lucas_Henrard_Lambin}. For an isotropic
coating $\varepsilon_\|=\varepsilon_\perp$ the polarizability reduces
to the standard result given in for instance Ref. 
\onlinecite{Bohren_Huffman}. The imaginary part of $\alpha_{vv}$
is directly accessible through experimental extinction or absorption
measurements. Explicitly, the extinction cross-section is
$\sigma_{vv}=\w \Im [\alpha_{vv}]$, where $\Im[\cdot]$ denotes the
imaginary part of the entity within the square brackets []. 
In order to obtain the response of the coated metallic
nanoparticle we must proceed by incorporating realistic microscopic
dielectric functions for the coating and metal into the expression
(\ref{eq:alpha1}) for the dipolar particle polarizability.

Let us consider the interior metallic region. In a standard fashion we
assume that the dielectric function of the metal is described by a
Drude function:
   \be
\varepsilon_{\rm in}(\w )=\varepsilon_\infty -\frac{\w_p^2}{\w (\w +i\Gamma_p )}
\label{eq:epsilon_i}  
  \ee 
with $\w_p$ being the plasmon frequency, $\Gamma_p$ is a
phenomenological damping parameter and $\varepsilon_\infty$ is the
dielectric function for large frequencies. We take
$\varepsilon_\infty=1$ throughout this study. An {\em uncoated}
metallic ellipsoidal particle in vacuum ($\epsilon_m=1$) has dipolar
plasmon resonance frequencies
  \be
\w_{pv}=\w_p \sqrt{n_v^i},\label{eq:plasmon_freq}
  \ee
where $n_v^i$ is the (purely geometric) depolarization factor of the
inner ellipsoidal surface
\cite{Bohren_Huffman,Landau_Lifshitz_ecm,Ambjornsson_Mukhopadhyay}
(see also appendix \ref{sec:appendixA}).

We now give the dielectric function of the coating; we assume that the
coating consists of molecules with dipolar polarizabilities which are
diagonal but with different components in normal and tangential
directions; we denote the polarizabilities by $\gamma_{\|}$
($\gamma_{\perp}$), {[}where a subscript $\perp$ ($\|$) denotes
polarizability component perpendicular (parallel) to the metallic
surface normal{]}. Let us relate $\gamma_{\|}$ and
$\gamma_{\perp}$ to the dielectric functions $\ecpp(\w)$ and
$\ecpr(\w)$ appearing in coated ellipsoid polarizability
$\alpha_{vv}$: assuming that the surface is locally flat, and imposing
the conditions that the normal component of the total macroscopic
electric field and the tangential component of the displacements field
are continuous across the surface separating the molecular coating
from the surrounding, we straightforwardly arrive at the following
relation between the molecular polarizabilities and the coating
dielectric functions \cite{Bagchi_Barrera_Fuchs}
 \bea
\varepsilon_{\perp}(\w)&=&1+\frac{4\pi\gamma_\perp(\w )}{v_0},
\nonumber\\ 
\varepsilon_{\|}(\w )^{-1}&=&1-\frac{4\pi\gamma_\|(\w )}{v_0},\label{eq:P}
 \eea 
where $v_0$ is the unit cell volume per
molecule.\cite{Ambjornsson_Apell_Mukhopadhyay} Explicitly, we take the
following form for the renormalized molecular polarizabilities
 \be 
\gamma_{\mu}(\w )
 =\gamma_{\infty \mu}+(\gamma_{0\mu}-\gamma_{\infty \mu})
 \frac{\w_{0\mu}^2}{\w_{0\mu}^2-\w^2-i\w \Gamma_\mu}\label{eq:pol_single_mol} 
\ee
where $\gamma_{0\mu}$ is the static polarizability of the molecules,
$\gamma_{\infty \mu}$ is the high frequency polarizability,
$\w_{0\mu}$ is a resonance frequency and $\Gamma_\mu$ is a damping
parameter.\cite{Ambjornsson_Apell,Bagchi_Barrera_Fuchs} It is
sometimes convenient to characterize the electromagnetic response
properties of the coating by the effective oscillator strength
(compare to Ref. \onlinecite{Wiederrecht_Wurtz_etal}) $c_{0\mu}=4\pi
(\gamma_{0\mu}-\gamma_{\infty\mu})/v_0$, and the large frequency
refractive indices ${\rm n}_\perp={\varepsilon_{c\perp}(\w
  =\infty)}^{1/2}=(1+4\pi \gamma_{\infty\mu}/v_0)^{1/2}$ and ${\rm
  n}_\|={\varepsilon_{c\|}(\w
  =\infty)}^{1/2}=(1-4\pi\gamma_{\infty\mu}/v_0)^{-1/2}$.  We point
out that the induced dipole coupling between molecules (and image
dipoles in the metal) in general, introduce extra anisotropy by
renormalizing the resonance frequencies, static polarizabilities,
large frequency polarizabilities and damping parameters compared to
their ``bare'' values,
\cite{Bagchi_Barrera_Fuchs,Ambjornsson_Apell,Ambjornsson_Apell_Mukhopadhyay}
and all such anisotropies are included in the expression
(\ref{eq:pol_single_mol}).

To summarize, the general procedure for obtaining the polarizability
$\alpha_{vv}(\w)$ ($v=x,y,z$) for an ellipsoidal metallic nanoparticle
coated with an anisotropic molecular layer is: Consider a nanoparticle
with principal semiaxes $a_x$, $a_y$, and $a_z$. The size of the
molecules together with these nanoparticle principal semiaxes
determine the outer principal semiaxes $b_x$, $b_y$ and $b_z$, see
Fig. \ref{fig:particle}. From these six principal semiaxes one
calculates the the total volume $V_o$, and the shape parameters $s$,
$e_i^2$, $e_o^2$ (for small coating thickness $e_i^2$ and $e_o^2$ are
related through the relative coating thickness $\Delta_V$). The
problem is completed by specifying the metallic nanoparticle parameters in
Eq. (\ref{eq:epsilon_i}), and the renormalized polarizability
parameters of Eq. (\ref{eq:pol_single_mol}); alternatively, one
may use experimental results for the nanoparticle and coating
dielectric functions. Finally, the parameters above are used in the
expression for $\alpha_{vv}(\w)$ explicitly given in Eq.
(\ref{eq:alpha1}).

For the case where the coating resonance frequency $\w_{0\mu}$ is
close to being degenerate with one of the particle plasmon resonances
(i.e., $\w_{pv}$ of Eq. (\ref{eq:plasmon_freq})) one expect
hybridization between the two resonances.\cite{HalasScience} In order
to address this point we proceed by finding an approximate expression
for the resonance frequencies for $\alpha_{vv}(\w)$. Assuming a thin
coating $\Delta_V\ll 1$, and utilizing the near resonance
approximation (the large $u$-result
\cite{Ambjornsson_Mukhopadhyay,Ambjornsson_Apell_Mukhopadhyay}) we
obtain a simplified form for $\alpha_{vv}$, as detailed in appendix
\ref{sec:appendixA}. Below we use this $\alpha_{vv}$ for obtaining
approximate expressions for the coupled (hybridized) resonance
frequencies and for the extinction peak height at these
resonances. The damping constants associated with the hybridized
resonances are given in Eq. (\ref{eq:Gamma_final}). We point out the
the approximate expressions for $\alpha_{vv}$ given in appendix
\ref{sec:appendixA} are useful for investigating other quantities as
well.

The resonance frequencies are obtained by finding the poles of
$\alpha_{vv}$; using the results in the appendix we find that
resonances occur at frequencies $\w=\w_{v\mu}^{\pm}$ given by (see
Eqs. (\ref{eq:Rabi_paral}) (\ref{eq:D_paral_app}),
(\ref{eq:Rabi_perp}) and (\ref{eq:D_perp_app}))
  \be
(\w_{v\mu}^{\pm})^2=\w_{a\mu}^2 \pm \left( \w_{d\mu}^4 
+\w_{0\mu}^4 D_{v\mu} ^2\right)^{1/2},\label{eq:Rabi}   
  \ee
where for brevity, we have introduced the notations 
$\w_{a\mu}^2 =[(\w^o_{pv})^2 +\w_{0\mu}^2]/2$ and 
$\w_{d\mu}^2=[(\w^o_{pv})^2 -\w_{0\mu}^2]/2$, with 
$\w^o_{pv}=\w_p\sqrt{n_v^o}$; 
we have also introduced above a (dimensionless) parallel coupling strength 
  \be
 D_{v\|}=\left[ \Delta_V g_{v\|}c_{0\|} (1-n_v^o)
 \frac{\w_p^2 -\w_{a\|}^2}{\w_{0\|}^2} \right]^{1/2}, \label{eq:D_paral}
  \ee
and a perpendicular coupling strength
  \be
 D_{v\perp}=\left[\Delta_V g_{v\perp}c_{0\perp}n_v^o 
 \frac{\w_{a\perp}^2}{\w_{0\perp}^2} \right]^{1/2}.\label{eq:D_perp}
  \ee
with the effective oscillator strength $c_{0\mu}= 4\pi
(\gamma_{0\mu}-\gamma_{\infty\mu})/v_0$ as defined before, and the geometric
factors defined as
  \be
g_{v\perp}=1-\frac{1}{F_of_v^o},\quad {\rm and}\quad 
g_{v\|}=\frac{1}{F_o f_v^o}.\label{eq:osc_strengths}
  \ee
where $f_v^o=f_v(s;e_o^2)$ so that, $f_x^o=1$, $f_y^o=1-se_o^2$, $f_z^o=1-e_o^2$
and $F_o=1+(1-se_o^2)^{-1}+(1-e_o^2)^{-1}$. The quantities
$g_{v\mu}$ appear also in the polarizability for an anisotropic shell
\cite{Ambjornsson_Apell_Mukhopadhyay} and satisfies the three sum
rules $g_{x\|}+g_{y\|}+g_{z\|}=1$,
$g_{x\perp}+g_{y\perp}+g_{z\perp}=2$ and $g_{v\perp}+g_{v\|}=1$.
Eq. (\ref{eq:Rabi}) has the same form as that for two dipole coupled
oscillators, see for instance Ref. \onlinecite{Pippard} (also compare
with Ref. \onlinecite{HalasScience}). The strength of the
molecular-plasmon interaction is characterized by the polarization and
molecular orientation dependent coupling strength $D_{v\mu}$, which depend
on the relative coating volume parameter $\Delta_V$, on the molecular 
response properties through $c_{0\mu}$, as well as on different geometric quantities. 

For a {\em sphere} ($e_o^2\rightarrow 0$) we have that $g_{v\|}=1/3$,
$g_{v\perp}=2/3$, and $n_v^o=1/3$ and Eqs. (\ref{eq:D_paral}) and
(\ref{eq:D_perp}) become
  \bea
D_{v\|} |_{\rm sph}&=&\left[\Delta_V \frac{c_{0\|}}{27}
\left(5\frac{\w_p^2}{\w_{0\|}^2} -3\right) \right]^{1/2}, \nonumber\\
D_{v\perp} |_{\rm sph}&=&\left[\Delta_V \frac{c_{0\perp}}{27}
\left(\frac{\w_p^2}{\w_{0\perp}^2} +3 \right)\right]^{1/2}. \label{eq:D_sphere}
  \eea
The equations above quantify the degree of hybridization and its
dependence on molecular orientation for a coated metallic sphere.  
\cite{cylindrical_case}

We notice from Eq. (\ref{eq:Rabi}) that if the coating and particle
plasmon resonance frequencies are well separated, in the sense $
D_{v\mu}^2 \ll \w_{d\mu}^4/\w_{0\mu}^4$, then the resonance frequencies are the
uncoupled ones: $\w_{pv}$ and $\w_{0\mu}$.

In contrast, for $D_{v\mu}^2\gg \w_{d\mu}^4/\w_{0\mu}^4$, we have resonances
approximately at
\be
\w_{v\mu}^{\pm}\approx \w^o_{pv} (1\pm D_{v\mu})^{1/2} \quad{\rm for}\quad \w^o_{pv}=\w_{0\mu},
\ee
and 
\bea
D_{v\|} & \approx &(1-n_v^o) (\Delta_V g_{v\|}c_{0\|}/n_v^o )^{1/2},\nonumber\\
D_{v\perp} & \approx &(\Delta_V g_{v\perp} c_{0\perp}n_v^o )^{1/2} .\label{eq:D_at_res}
\eea
Thus, for $\w^o_{pv}\approx \w_{0\mu}$, we get $D_{v\mu} \propto
\sqrt{\Delta_V}$ and we have strong hybridization between the
nanoparticle and coating resonance frequencies, as will be illustrated
in more detail in the next section. We also notice from
Eq. (\ref{eq:D_at_res}) that for small $n_v^o$, $D_{v\|}$ can be
substantially larger than $D_{v\perp}$, which can give a wider
separation between $\w_{v\|}^{+}$ and $\w_{v\|}^{-}$ compared with
that for $\w_{v\perp}^{\pm}$.

The extinction peak height at one of the resonance frequency
$\w_{v\mu}^{\pm}$ is $\sigma_{vv} (\w=\w_{v\mu}^{\pm})$; using
Eq. (\ref{eq:T_v_def}) we convert this quantity into a scaled peak
height $T^{\pm}_{v\mu}=4\pi\sigma_{vv}(\w^{\pm}_{v\mu})/(V_o 
(\w^{\pm}_{v\mu})^2)$ at the resonances. Employing the thin coating
approximation and assuming that we are close to a resonance
Eqs. (\ref{eq:T_v_paral}) and (\ref{eq:T_v_perp}) apply, i.e. we have
the approximate result
 \bea
T^\pm_{v\mu}&=&\frac{\w_p^2}{(\w_{v\mu}^{\pm})^2}\nonumber\\
& &\hspace{-1cm}\times\frac{\big(\w_{0\mu}^2-(\w_{v\mu}^{\pm}\big)^2)}
{\Gamma_p\big(\w_{0\mu}^2-(\w_{v\mu}^{\pm})^2\big)+
\Gamma_{\mu}\big((\w^o_{pv})^2-(\w_{v\mu}^{\pm})^2\big)}\label{eq:T_v_general},
  \eea
which for $\w^o_{pv}\approx \w_{0\mu}$ gives 
 \be
T^\pm_{v\mu}=\frac{1}{n_v^o (\Gamma_p+\Gamma_{\mu})}\,\frac{1}{1\pm D_{v\mu}},\label{eq:T_v_hybrid}
 \ee
with $D_{v\mu}$ given by Eq. (\ref{eq:D_at_res}).

For the case of an uncoated metallic nanoparticle the scaled peak height
takes the form $T_{vp}=1/(n_v^o \Gamma_p)$ (see Eq. (\ref{eq:T_v_plasmon})).
Eq. (\ref{eq:T_v_hybrid}) shows that at strong coupling (i.e. for
$\w^o_{pv}\approx \w_{0\mu}$), $T^\pm_{v\mu}$ are of the same order
of magnitude as $T_{vp}$.

For a very thin coating without any interior metal we have: 
$T_{v\mu}(\varepsilon_i=1)\approx \Delta_V g_{v\mu} c_{0\mu}/\Gamma_{\mu}$
(see Eqs. (\ref{eq:alpha_paral}) and (\ref{eq:alpha_perp}), and
Ref. \onlinecite{Ambjornsson_Apell_Mukhopadhyay}), which is much smaller
than $T_{vp}$. Comparing with Eq. (\ref{eq:T_v_general}) we notice that
$T^\pm_{v\mu}/T_{v\mu}(\varepsilon_i=1)\propto (\Delta_V)^{-1}$ (note
however that $\w_{v\mu}^{\pm}$ depends on $\Delta_V$ ),  revealing
substantial surface enhancement (see next section).

\section{Results and Discussions.} 

In the following, we illustrate the electromagnetic response
properties of a coated metallic nanoparticle via the extinction
cross-section $\sigma_{vv}=\w \Im[\alpha_{vv}]$ for the case of a
prolate ($a_y=a_z$, i.e.  $s=1$, ``cigar-shaped'') spheroid. We
consider two different situations: the case when the molecules in the
coating have their ``resonant'' axes (i) parallel and (ii)
perpendicular to the surface normal. The polarizability component
along the resonant axis is $\alpha_\|$ in case (i) and
$\alpha_\perp=\alpha_\|/2$ in case (ii), which corresponds to coatings
with identical but differently oriented molecules. For each of the two
cases, the external field $\vec{E}_0$ is applied either along the long
axis ($x$-axis) or the short axis ($y$- or $z$-axis).

In Fig. \ref{fig:particle}b, the polarization averaged extinction
cross-section for an {\em uncoated} metallic spheroid is shown
together with the cross-section for an spheroidally shaped molecular
layer {\em without} an interior nanoparticle and with the resonant
transition dipole moments oriented in the parallel and perpendicular
directions. There are two plasmon peaks, as expected for a spheroidal
metal particle (two of the principal axes equal), with resonance
wavelengths to the red ($\vec{E}_0 \| x$-axis) and to the blue
($\vec{E}_0 \| y,z$-axes) of the familiar LSP resonance of a sphere
($\lambda/\lambda_p = \sqrt{3}\approx 1.73$). The extinction peaks of the
molecular layer is about 50 times weaker than the plasmons for both
molecular orientations. Notice that, for geometric reasons, the
extinction peak height for the molecular layer is smallest for the
case where the external field is in the $x$-direction, with the
molecules oriented in the parallel direction; see
Ref. \onlinecite{Ambjornsson_Apell_Mukhopadhyay} for a thorough discussion
on geometric and molecular orientation dependent effects in the
response of an ellipsoidal molecular layer. Also notice the small
difference in peak position between case (i) and (ii); in Ref.
\onlinecite{Ambjornsson_Apell_Mukhopadhyay} it was shown that for
$\Delta_V\ll 1$ there are no curvature induced shifts for the
molecular layer resonance frequencies (see also appendix
\ref{sec:appendixA}). Here, $\Delta_V$ is not sufficiently small and
there are some minor shifts of the resonance frequencies.

\begin{figure}
  \scalebox{0.53}{\includegraphics{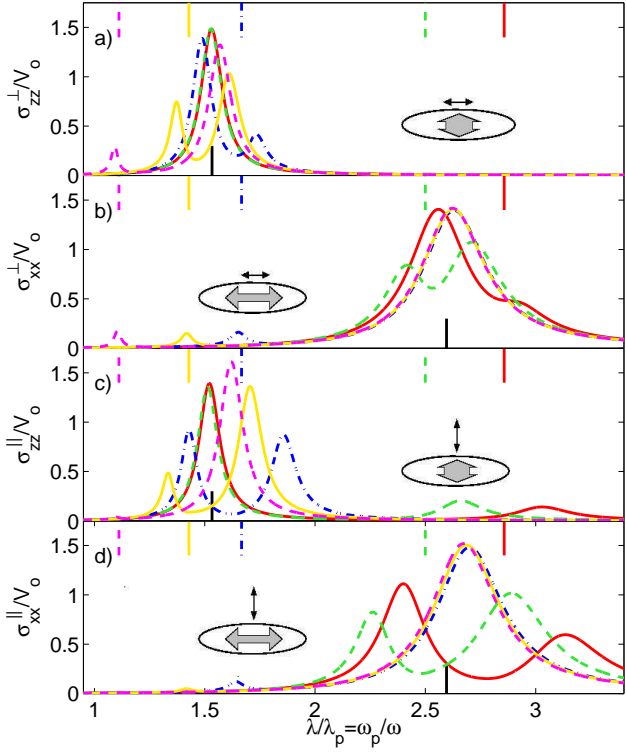}}
  \caption{Extinction spectrum for a prolate coated
 spheroid with different orientations of the molecules on the
 surface. The spectrum is in units of the total volume $V_o$ of the
 coated spheroid. $\sigma^\mu_{vv}=\w \Im [\alpha_{vv}]$,
 ($\mu=\|,\perp$ and $v=x,z$) and $\lambda=2\pi c/\w$ is the wavelength
 of the external electromagnetic field.  a) The external field is
 along the $z$-axis (short axis) of the spheroid. The molecules have
 their resonant axes {\em perpendicular} to the metallic particle normal;
 b) the external field along the $x$-axis (long axis), and the
 molecules are {\em perpendicular} to the layer normal; c) the field is
 along the $z$-axis, and the molecules have their resonant axis {\em
 parallel} to the layer normal; d) the field is along the
 $x$-axis, and the molecules are oriented {\em parallel} to the
 layer normal. The different colors are spectrum for different
 molecular resonant wavelengths; the upper vertical bars indicate
 these resonance wavelengths $\lambda_{0\mu}=2\pi c/\w_{0\mu}$. Here, we
 chose the following values for $\w_{0\mu}$: $0.35\w_p$, $0.4\w_p$,
 $0.6\w_p$, $0.7\w_p$, $0.9\w_p$.  The lower vertical bar is the
 particle plasmon resonance wavelengths $\lambda_{pv}=2\pi c/\w_{pv}$
 ($v=x,z$). The same parameters as in Fig. \ref{fig:particle} were
 used. Notice the strong hybridization when $\lambda_{0\mu}\approx
 \lambda_{pv}$ and increased peak height (surface enhanced absorption)
 for the molecular resonance, compared with Fig. \ref{fig:particle}b.}
  \label{fig:absorption}
\end{figure}

In Fig. \ref{fig:absorption}, we show extinction spectra
corresponding to the full coated spheroid response problem for the
four distinct cases mentioned above, i.e molecular orientation
perpendicular (a, b) or parallel (c, d) to the surface normal and
incident field parallel to the short (a, c) or the long (b, d)
spheroid axes. The different colors of the curves corresponds to
different coating resonance frequencies $\w_{0\mu}$. It is clear that
there is strong hybridization between the coating and plasmon
resonances whenever $\w_{0\mu}\approx \w_{pv}$ ($\mu=\|,\perp$ and
$v=x,y,z$) in all cases. However, the degree of hybridization differs
greatly between the different molecular orientations and polarization
configurations.

\begin{figure}
  \scalebox{0.45}{\includegraphics{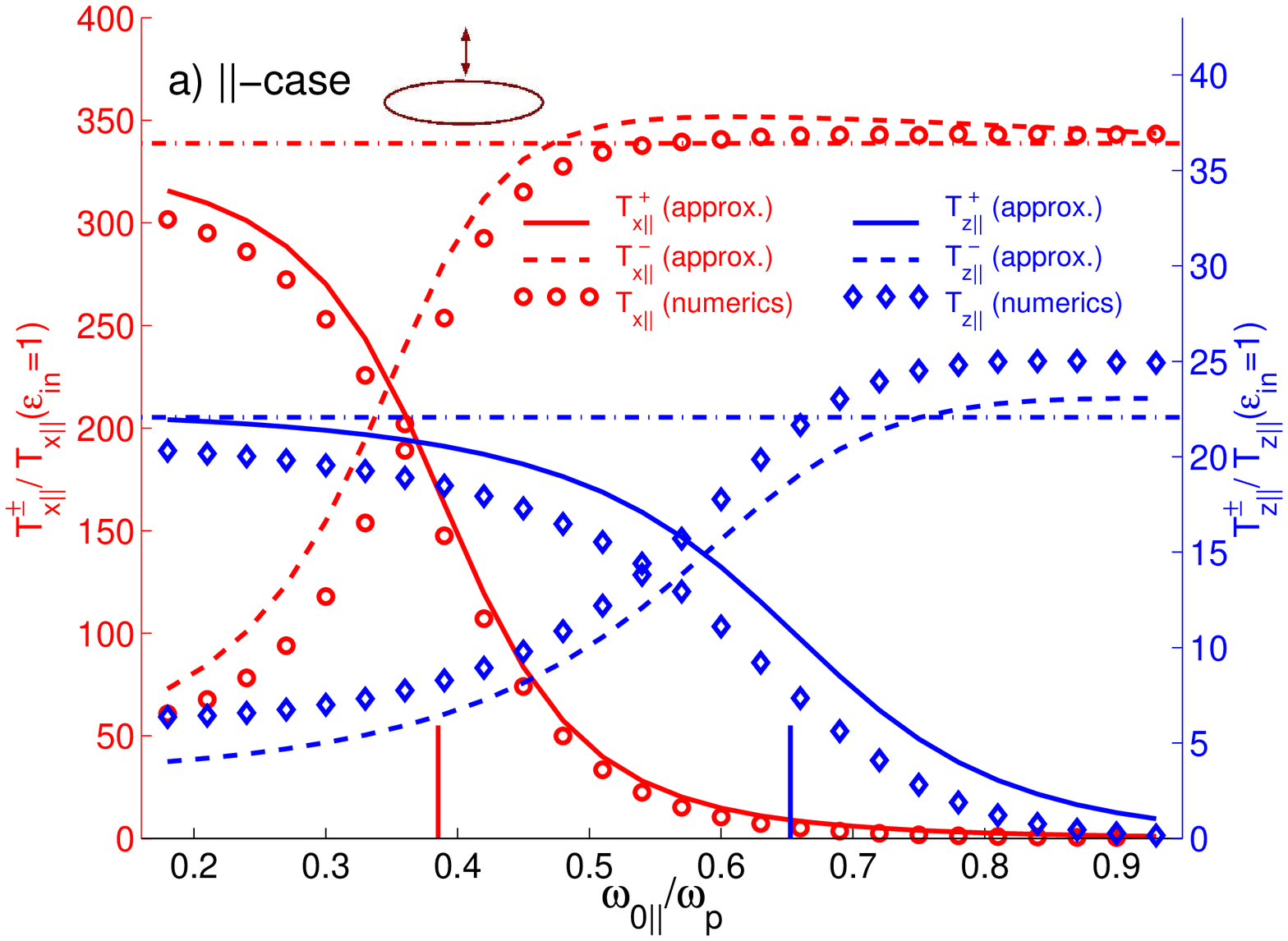}}
  \scalebox{0.45}{\includegraphics{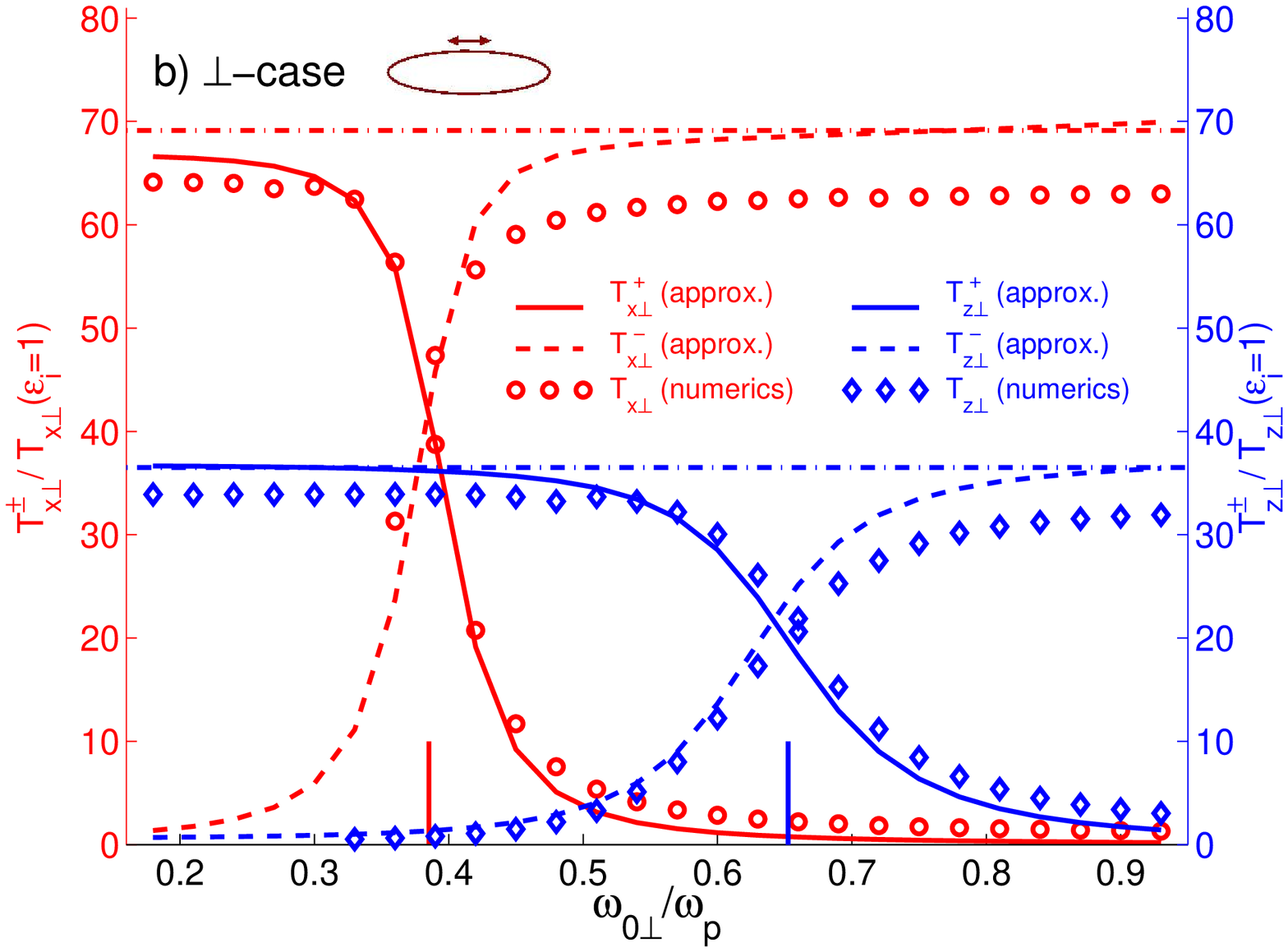}}
  \caption{ Scaled peak height for the coupled
    resonances in a coated prolate metal spheroid for the case a) when
    the coating molecules have their resonant axis {\em parallel} to
    the surface normal and b) for a {\em perpendicular} resonance. The
    resonance frequencies $\w^{\pm}_{v\mu}$ of the extinction (see
    Figs. \ref{fig:absorption} and \ref{fig:absorption_2d}) for
    different $\w_{0\mu}$ was numerically determined and used to
    calculate the scaled peak height
    $T^{\pm}_{v\mu}=4\pi\sigma_{vv}(\w^{\pm}_{v\mu})/(V_o
    (\w^{\pm}_{v\mu})^2)$ ($v=x,z$ and $\mu=\|,\perp$), see
    Eq. (\ref{eq:T_v_def}). The peak heights were normalized by the
    molecular layer peak heights $T_{v\mu}(\varepsilon_i=1)$ obtained
    from Fig. \ref{fig:particle}. The vertical solid lines correspond
    to the two plasmon resonances, $\w_{px}$ and $\w_{pz}$,
    Eq. (\ref{eq:plasmon_freq}). The dashdotted horizontal lines
    correspond to the plasmon peak heights as given in the text right
    after Eq. (\ref{eq:T_v_general}). The solid and dashed curves are
    the approximate results given in Eq. (\ref{eq:T_v_general}). The
    same parameters as in Fig.  \ref{fig:particle} were used. Notice
    the different ranges of the ordinates, and that the abscissas are in
    frequency units.}
 \label{fig:T_v}
\end{figure}

In order to investigate the peak height of the resonances (see
Fig. \ref{fig:absorption}) Fig. \ref{fig:T_v} show the scaled peak
height $T^\pm_{v\mu}$ (defined in Eq. (\ref{eq:T_v_def})) compared to
the scaled peak height $T_{v\mu}(\varepsilon_i=1)$ for a molecular
layer (obtained from Fig. \ref{fig:particle}), for different coating
resonance frequencies.  A comparison to the result given in
Eq. (\ref{eq:T_v_general}) is also made. We notice the two coupled
resonances obtain similar peak height as $\w_{0\mu}$ approaches the
particle plasmon frequency $\w_{pv}$. We also point out that the
approximate expression Eq. (\ref{eq:T_v_general}) works surprisingly
well, considering that it is based on a close to resonance and thin
coating approximation, even far off the resonances.

When the molecular resonance is far above or far below a plasmon
resonance, we have a situation that can best be described as
surface-enhanced absorption from the molecular layer
\cite{Moskovits1985}, i.e the magnitude of the molecular resonance
peak is greatly enhanced but its position and width is not changed
dramatically from the case of a "free" molecular layer, see Figs
\ref{fig:absorption} and \ref{fig:T_v}. In this regime, the particle
plasmons (at $\lambda_{pv}$, with $v=x,y,z$) become either red-shifted
or blue-shifted relative to the uncoated metallic spheroid resonance
wavelength depending on the molecular resonance wavelength
$\lambda_{0}$. Interestingly, the enhanced absorption from the
molecular layer is not symmetric on the two sides of the plasmon. This
can be seen when comparing Fig. 2 a) and c) (see also
Fig. \ref{fig:T_v}), corresponding to polarization along the short
axis of the spheroid. When the molecular resonance is far to the red
of the plasmon, absorption enhancement is seen for the case when the
molecular resonance axis is parallel to the surface normal (Fig.  2
c), but not for the perpendicular case (Fig. 2 a). The situation is
reversed (although less pronounced) when the molecular resonance is to
the blue of the plasmon, i.e absorption enhancement occurs for the
perpendicular but not for the parallel case. Similar effects are seen
in Fig. 2 b) and d), corresponding to polarization along the long
axis. From Fig. \ref{fig:T_v}a) we notice that particularly strong
surface enhancement (roughly a factor 50 for the extinction cross
section for the parameters used here) occur to the red of the plasmon
for the parallel case with the incident field along the x-axis.

The differences above can be understood as follows: To the red of a
plasmon resonance, the induced field from the particle is in phase
with the applied field and normal to the surface at the poles of the
particle (where the poles are defined by the direction of the induced
dipole), resulting in enhanced absorption from molecules oriented
parallel the surface normal. Molecules with the perpendicular
orientation instead couple mainly to the total field around the
equator of the particle, where the field is perpendicular to the
surface normal. But to the red this field is weak, because here the
induced field is out-of-phase with the applied field. Hence, there is
very little enhanced absorption to the red of the plasmon in Fig 2 a)
but a large enhancement in Fig. 2 c). The same type of arguments apply
to the case when the molecular resonance is to the blue of the
plasmon, but here the situation becomes reversed because the induced
particle dipole is out-of-phase with the incident field.\cite{Kottman}
\begin{figure}
 \scalebox{0.45}{\includegraphics{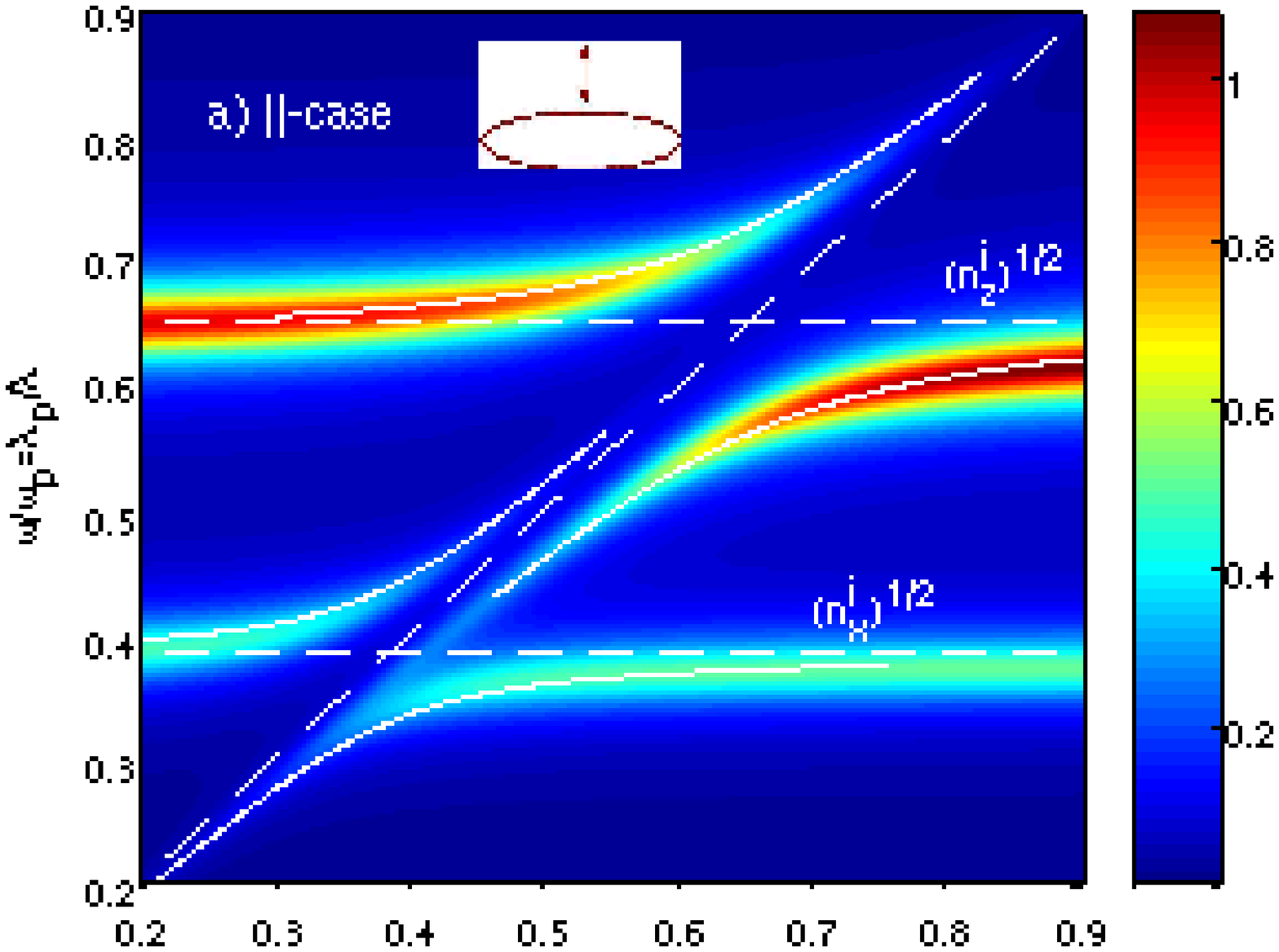}}
 \scalebox{0.45}{\includegraphics{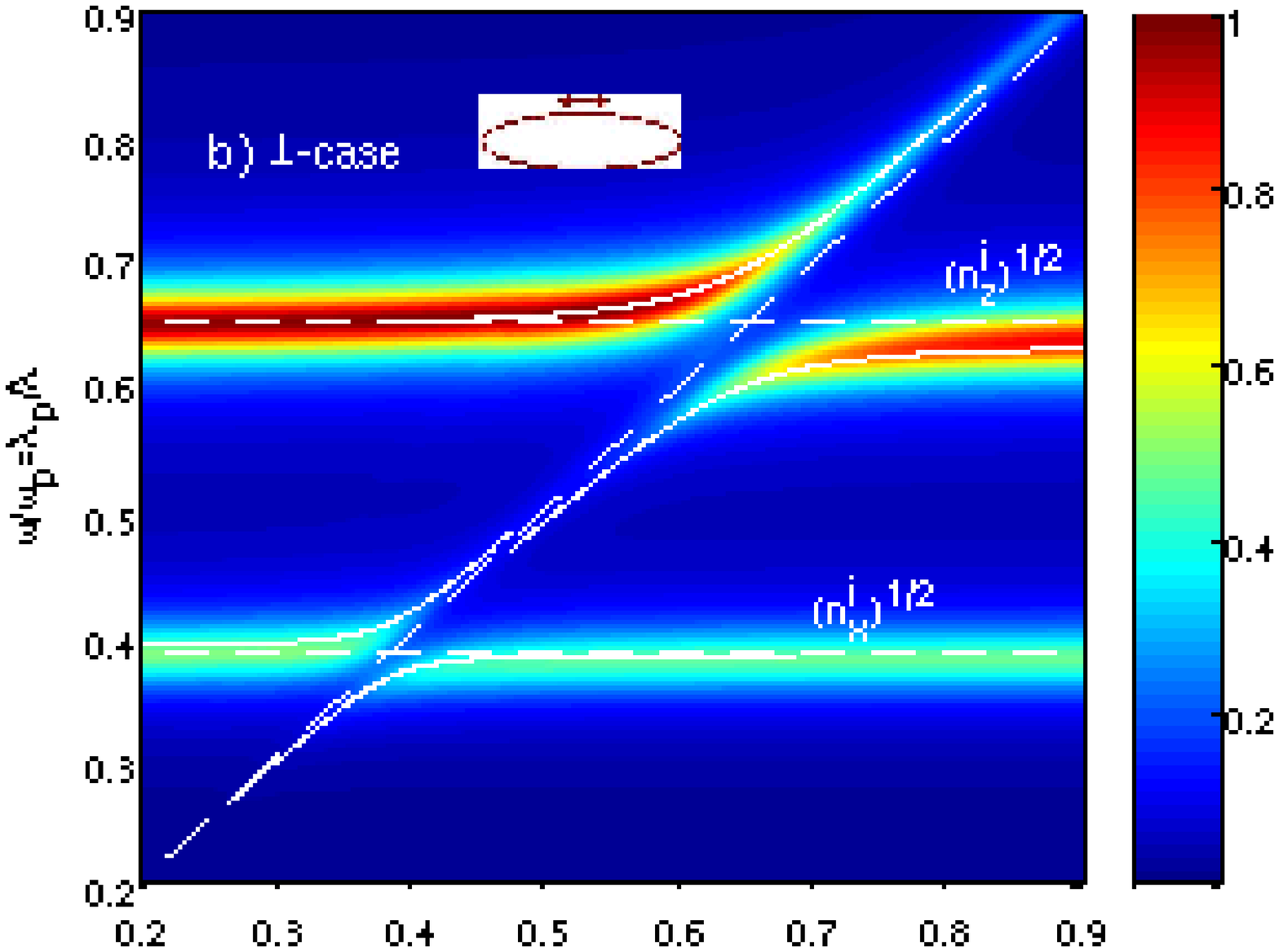}}
  \caption{Illustration of strong coupling and "avoided
   crossing" between the plasmonic and molecular resonances of a
   coated prolate metal spheroid for the case a) when the molecules
   have their resonant axis {\em parallel} to the surface normal and
   b) for a {\em perpendicular} resonance. The horizontal dashed lines
   correspond to the two plasmon resonances, $\w_{px}$ and
   $\w_{pz}$, and the diagonal dashed lines denotes
   $\w=\w_{0\mu}$. The solid white curves are the approximate results
   for the coupled resonances as given in Eq.
   (\ref{eq:Rabi}). The same parameters as in Fig.
   \ref{fig:particle} were used. Notice that the axes are in
   frequency units.}
 \label{fig:absorption_2d}
\end{figure} 

The case when the molecular resonance is degenerate with one of the
plasmons corresponds to a regime that can not be described in terms of
enhanced absorption. Instead, we have two completely hybridized
resonances that exhibit "avoided crossing", analogous to the case of
two strongly coupled (quantum or classical) harmonic oscillators, as
is seen in Eq. (\ref{eq:Rabi}).  Fig. \ref{fig:absorption_2d}
illustrates this behavior for the two molecular orientations
considered. In the vocabulary of Ref. \onlinecite{HalasScience}, the
high frequency modes $\w_{v\mu}^{+}$ in Eq. (\ref{eq:Rabi})
correspond to "anti-bonding" combinations, i.e. the case where the
induced dipole on the particle is out-of-phase with the molecular
transition dipoles, while the "bonding" modes $\w_{v\mu}^{-}$
corresponds to the case when the excitations are in phase.  In a
diabatic representation, this corresponds to Rabi oscillations in
which the excitation energy oscillates back and forth between the
plasmonic particle and the molecular layer. As can be seen in
Fig. \ref{fig:absorption_2d}, the mode splitting is much more
significant when the molecular resonance is oriented parallel to the
surface normal (Fig. \ref{fig:absorption_2d}a) than for the
perpendicular case (Fig. \ref{fig:absorption_2d}b). This difference
simply reflects the predominant polarization of the local field at the
surface (the field would be strictly parallel to the surface normal
for a perfect conductor at zero frequency).  We also note that the
splitting (expressed in frequency units) is larger for the plasmon
that corresponds to polarization along the short axis of the prolate
spheroid for both molecular orientations, which is somewhat surprising
considering that the field enhancement is highest at the sharp ends of
the spheroid.  \cite{Kottman} However, the coupling strength is
determined by the surface integrated local field at the two plasmon
resonance frequencies, and for geometric reasons this quantity is
higher for the doubly-degenerate short-axis plasmon in the present
case.

\begin{figure}
 \scalebox{0.45}{\includegraphics{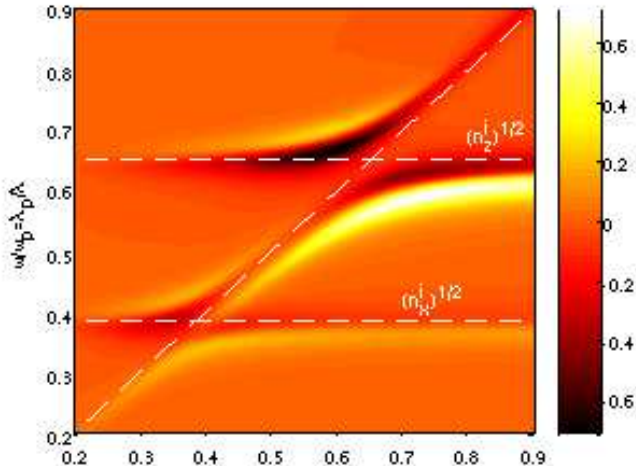}} 
 \caption{ Difference in averaged extinction
   cross-section $\Delta \sigma/V_o =(\sigma^\|- \sigma^\perp)/V_o$
   (indicated by the different colors) between the parallel and
   perpendicular cases as shown in Fig. \ref{fig:absorption_2d} a)
   and b), as a function molecular resonance frequency
   $\w_0=\w_{0\|}=\w_{0\perp}$ and external field frequency $\w$. Note
   that the axes are in frequency units.}
 \label{fig:diff_spectrum}
\end{figure}

To the best of our knowledge, there are no experimental studies that
have directly probed the orientation dependent plasmon-molecule
coupling discussed here. However, the advanced nanofabrication and
molecular functionalization technologies of today clearly make such
studies a realistic possibility. One of the most interesting options
could be to use a plasmonic nanoparticle covered by an ordered layer
of chromophores as a nanoscopic bio/chemo sensor. For example, it can
be expected that the exact orientation of the chromophores on the
surface will be sensitive to pH and to the interaction with molecules
in solution. The chromophores might also be incorporated as functional
groups into larger biomacromolecules that change
orientation/conformation upon biorecognition reaction with an
analyte. Such sensing reactions would then affect the coupling between
the surface plasmons and the chromophores, giving rice to pronounced
changes in the extinction or scattering spectrum of the composite
nanoparticle. To illustrate this possibility, Fig.
\ref{fig:diff_spectrum} shows the difference spectra between the
parallel and perpendicular orientations discussed above. As expected,
the difference is largest for molecular resonance wavelengths that
overlap the LSP modes. However, a noticeable difference is observed
even far off the plasmon resonances, indicating that molecular
orientation effects needs to be taken into account also in classical
nanoparticle refractive index sensing experiments.

\section{Summary} 
We have described an analytic method for calculating the optical
response of arbitrary ellipsoidal nanoparticles with anisotropic
molecular coatings in the small particle limit and applied this theory
to the case of a prolate metal spheroid covered with chromophores. The
results show that the hybridization between the molecular resonance
and the localized plasmon resonances of the nanoparticle is highly
anisotropic. It is suggested that this sensitivity can be utilized for
novel bio/chemo sensing applications that are based on molecular
orientation rather than refractive index contrast.

\begin{acknowledgments} 
M.K. acknowledges stimulating discussions with Peter Nordlander and a
Texas Instruments Visiting Professorship that supported part of this
work. G.M. acknowledges the hospitality of the Department of Applied
Physics, Chalmers University. This work was financially supported by
the Swedish Research Council.
\end{acknowledgments} 

\appendix

\section{Approximate expression for coupled resonances and peak
heights and widths}\label{sec:appendixA}

In this appendix we use the large $u$ (close to a resonance) and thin
coating approximations for the coated ellipsoid polarizability given
in Eq. (\ref{eq:alpha1}) in order to derive approximate expressions
for the coupled resonance frequencies as well as for the peak heights
and widths of these excitations.

Close to a resonance ($u$ large) and for a thin coating ($\Delta_V\ll
1$) we have that Eq. (\ref{eq:alpha1}) becomes
  \be   
\alpha_{vv}=\frac{V_o}{4\pi
n_v^o}\frac{J_v(\kappa=-1)}{J_v(\kappa=1/n_v^o-1)}\label{eq:alpha_app},
  \ee
with
  \be
J_v(\kappa)=\varepsilon_i+\kappa +\Delta_V g_{v\perp}\varepsilon_{\perp}+
\Delta_V \varepsilon_i g_{v\|}\varepsilon^{-1}_{\|}\kappa\label{eq:thin_coat_approx}
  \ee
in an identical fashion as the derivation given in Ref.
\onlinecite{Ambjornsson_Apell_Mukhopadhyay}. The geometric factors
$g_{v\|}$ and $g_{v\perp}$ ($v=x,y,z$) are given in
Eq. (\ref{eq:osc_strengths}). Explicit expressions for the dielectric
functions $\varepsilon_i$, $\varepsilon_{\perp}$ and
$\varepsilon_{\|}$ appearing in Eq. (\ref{eq:thin_coat_approx}) are
given in Eqs. (\ref{eq:epsilon_i}) and (\ref{eq:P}). Below we
investigate $J_v(\kappa)$ in the limits of (1.) no coating $\Delta_V=
0$; (2.) external field frequency close to the parallel resonance
frequency, $\w\approx \w_{0\|}$; (3.) external field frequency close
to the perpendicular resonance frequency, $\w\approx \w_{0\perp}$. In
(4.) we derive expressions for the peak widths of the hybridized resonances.

\subsection{No coating, $\Delta_V= 0$}

For the case of no coating, $\Delta_V= 0$, we have that
$J_v(\kappa)=\varepsilon_i+\kappa$ and therefore the polarizability,
Eq. (\ref{eq:alpha_app}), can be written 
  \be
\alpha_{vv}(\w)=\frac{V_o}{4\pi n_v^o} \frac{K_v(\kappa=-1)}
{K_v(\kappa=1/n_v^o-1)},\label{eq:alpha_no_coat}
  \ee
where $K_v$ equals $J_v$ multiplied a by a factor independent of $\kappa$, and explicitly,
  \bea
\Re[K_v(\kappa)]&=&\w_p^2-\w^2(1+\kappa),\nonumber\\
\Im[K_v(\kappa)]&=&-\w \Gamma_p (1+\kappa),\label{eq:K_no_coat}
  \eea
with $\Re [\cdot]$ ($\Im [\cdot]$) denoting the real (imaginary) part
of the entity inside [].  The resonance criterion
$\Re[K_v(\kappa=1/n_v^o-1)]=0$ gives, together with
Eq. (\ref{eq:K_no_coat}), the uncoated metallic ellipsoid resonance
frequencies:
  \be
\w_{{\rm res},v}^2=(\w^o_{pv})^2=\w_p^2 n_v^o\label{eq:w_res_no_coat}
  \ee
in agreement with Eq. (\ref{eq:plasmon_freq}) as it should (note that when
$\Delta_V=0$ then $n_v^i=n_v^o$). At one of the resonances the
extinction peak height $\sigma_{vv}(\w_{{\rm res},v})=\w_{{\rm res},v} 
\Im[\alpha_{vv}(\w_{{\rm res},v})]$ is:
  \bea
\sigma_{vv}(\w_{{\rm res},v})&=&-\frac{V_o}{4\pi n_v^o} \frac{\w
\Re[K_v(\kappa=-1)]}{\Im [K_v(\kappa=1/n_v^o-1)]}|_{\w=\w_{{\rm res},v}}\nonumber\\
&=&\frac{V_o}{4\pi n_v^o} \Gamma_p^{-1} \w_{{\rm res},v}^2\label{eq:sigma_v_no_coat},
  \eea
where we have used Eq. (\ref{eq:K_no_coat}) and $\w_{{\rm res},v}$
is given in Eq. (\ref{eq:w_res_no_coat}). It is convenient to define a scaled
peak height
  \be
T_v\equiv \frac{4\pi}{V_o}\,\frac{\sigma_{vv}(\w_{{\rm res},v})}{\w_{{\rm res},v}^2}\label{eq:T_v_def},
  \ee
so that, using Eq. (\ref{eq:sigma_v_no_coat}), we have
  \be
T_v=T_{vp}=\frac{1}{n_v^o \Gamma_p}.\label{eq:T_v_plasmon}
  \ee
Thus, for an uncoated nanoparticle $T_v$ is a measure of the life time
(inverse damping) of the plasmon excitation. We will below use the
definition, Eq. (\ref{eq:T_v_def}), to compute approximate scaled
peak heights of the coated metallic ellipsoid excitations.

\subsection{$\w\approx \w_{0\|}$}

For the case $\w\approx \w_{0\|}$ (and $\w_{0\|}\neq \w_{0\perp}$) we
neglect the ``perpendicular'' term in Eq. (\ref{eq:thin_coat_approx}), so that
$J_v(\kappa)=\varepsilon_i+\kappa+\Delta_V \varepsilon_i g_{v\|}
\varepsilon^{-1}_{\|}\kappa$. Using the explicit expressions for the
dielectric functions described by Eqs. (\ref{eq:epsilon_i}), and (\ref{eq:P})
along with (\ref{eq:pol_single_mol}) we find
  \bea
  K_v (\kappa) &=&\left[\w(\w+i\Gamma_p)(\w^2-\w_{0\|}^2+i\w\Gamma_{\|})\right]
  \times J_v(\kappa)\nonumber\\
  &\approx&(\w^2+i\w\Gamma_p-\w_p^2)(\w^2-\w_{0\|}^2+i\w\Gamma_{\|})\nonumber\\
  & &+(\w^2+i\w\Gamma_p-\w_p^2)\Delta_V g_{v\|} c_{0\|}\w_{0\|}^2 \kappa\nonumber\\ 
  & &+(\w^2+i\w\Gamma_p)(\w^2-\w_{0\|}^2+i\w\Gamma_{\|})\kappa,
  \eea
Assuming that $\Delta_V\ll 1$, and neglecting terms of order $\Gamma^2$
($\Gamma=\Gamma_p,\Gamma_{\|}$), and $\Gamma \Delta_V$ (the neglected
term $\Delta_V g_{v\perp}\varepsilon_{\perp}$ is of the same order as
the terms neglected here) we find that the polarizability can be written
in the same form as in Eq. (\ref{eq:alpha_no_coat}) now with
  \bea
\Re[K_v(\kappa)]&=&(\w_{0\|}^2-\w^2)[\w_p^2-\w^2(1+\kappa)]\nonumber\\
&&- \Delta_V g_{v\|} c_{0\|}\w_{0\|}^2 (\w_p^2-\w^2)\kappa,\nonumber\\
\Im[K_v(\kappa)]&=&-\w\Big(\Gamma_p \big( (\w_{0\|}^2-\w^2)(1+\kappa)\nonumber\\
&&+\Gamma_{\|}\big((\w_p^2-\w^2(1+\kappa)\big)\Big)\label{eq:K_paral}.
  \eea
We notice that the second term in the expression for $\Re[K_v(\kappa)]$
cannot be discarded because the first term may be close to zero. In the
limit $\Delta_V,\Gamma_{\|}\rightarrow 0$ the polarizability reduces to
the uncoated result from the previous subsection.

In the limit $\w_p,\Gamma_p\rightarrow 0$
(i.e. $\varepsilon_i=\varepsilon_\infty=1$) we recover, to lowest
order in $\Delta_V$, the result in
Ref. \onlinecite{Ambjornsson_Apell_Mukhopadhyay}:
  \be
\alpha_{vv}(\w)=\Delta_V\,\frac{V_o }{4\pi} g_{v\|}c_{0\|}\,
\frac{\w_{0\|}^2}{\w_{0\|}^2-\w^2-i \w \Gamma_{\|}}\label{eq:alpha_paral}
  \ee
as it should.

Turning back to the general expression Eq. (\ref{eq:K_paral}) we
find that the resonance condition $\Re[K_v(\kappa=1/n_v^o-1)]=0$ gives
resonance frequencies:
 \bea
(\w_{v\|}^{\pm})^2&=&\frac{\w_p^2 n_v^o+\w_{0\|}^2}{2}\nonumber\\
& & \hspace{-0.5 cm}\pm \left[ \Big(\frac{\w_p^2 n_v^o-\w_{0\|}^2}{2}\Big) ^2 
+\w_{0\|}^4 D_{v\|}^2\right]^{1/2}\label{eq:Rabi_paral}   
  \eea
with a dimensionless coupling strength $D_{v\|}$ defined by
  \be
D_{v\|}^2=\Delta_V g_{v\|}c_{0\|} (1-n_v^o)
\Big(\frac{\w_p^2}{\w_{0\|}^2} -\frac{\w_p^2 n_v^o +\w_{0\|}^2}
{2 \w_{0\|}^2}\Big).\label{eq:D_paral_app}
  \ee

Using Eq. (\ref{eq:T_v_def}) we find that the scaled peak height
$T^+_{v\|}$ ($T^-_{v\|}$) of the resonance characterized by
resonance frequency $\w_{v\|}^+$ ($\w_{v\|}^-$) becomes (for
$\w_p\neq 0$)
  \bea
T^\pm_{v\|}&=&\frac{\w_p^2}{(\w_{v\|}^{\pm})^2}\nonumber\\
& &\hspace{-1.5cm}\times \frac{\big(\w_{0\|}^2-(\w_{v\|}^{\pm})^2\big)}
{\Gamma_p\big(\w_{0\|}^2-(\w_{v\|}^{\pm})^2\big)+\Gamma_{\|}
\big(\w_p^2n_v^o-(\w_{v\|}^{\pm})^2\big)},\label{eq:T_v_paral}
  \eea
where we have used the fact that $(\w_{v\|}^{\pm})^2$ is of the order
$(\Delta_V)^{1/2}$, and neglected terms of order $\Delta_V$.

We notice from Eq. (\ref{eq:Rabi_paral}) that if $D_{v\|}\ll |[\w_p^2
n_v^o-\w_{0\|}^2]/[2 \w_{0\|}^2]|$, then the resonance frequencies are the
uncoupled ones: $\w_p\sqrt{n_v^o}$ and $\w_{0\|}$.

If instead $D_{v\|}\gg |[\w_p^2 n_v^o-\w_{0\|}^2]/[2 \w_{0\|}^2]|$,  
(i.e., for $\w_p^2 n_v^o \approx \w_{0\|}^2$), 
 we get $(\w_{v\|}^{\pm})^2 \approx \w_p^2 n_v^o[1\pm D_{v\|}]$, and
 \be
 T^\pm_{v\|}= \frac{1}{n_v^o (\Gamma_p+\Gamma_{\|})}\,\frac{1}{1\pm D_{v\|}}
  \ee
with $D_{v\|}=(\Delta_V g_{v\|}c_{0\|}(1-n_v^o)^2/n_v^o)^{1/2}$, 
showing strong hybridization of the plasmon and coating resonances.

\subsection{$\w\approx \w_{0\perp}$}

For the case $\w\approx \w_{0\perp}$ (and $\w_{0\|}\neq \w_{0\perp}$)
we neglect the ``parallel'' term in Eq. (\ref{eq:thin_coat_approx}),
so that $J_v(\kappa)=\varepsilon_i+ +\kappa +\Delta_V
g_{v\perp}\varepsilon_{\perp}$.  Using the explicit expressions for
the dielectric functions in Eqs.  (\ref{eq:epsilon_i}) and
(\ref{eq:P}) along with (\ref{eq:pol_single_mol}) an identical
analysis as in the preceding subsection shows that polarizability can
be written in the same form as in Eq. (\ref{eq:alpha_no_coat}) with
  \bea
\Re[K_v(\kappa)]&=&(\w_{0\perp}^2-\w^2)\big(\w_p^2-\w^2(1+\kappa)\big)\nonumber\\
& &-\Delta_V g_{v\perp}c_{0\perp}\w_{0\perp}^2\w^2,\nonumber\\
\Im[K_v(\kappa)]&=&-\w\Big(\Gamma_p(\w_{0\perp}^2-\w^2)(1+\kappa)\nonumber\\
& &+\Gamma_{\perp}\big(\w_p^2-\w^2(1+\kappa)\big)\Big)\label{eq:K_perp}.
  \eea
In the limit $\Delta_V,\Gamma_{\perp}\rightarrow
0$ we recover the uncoated result.

In the limit $\w_p,\Gamma_p\rightarrow 0$
(i.e. $\varepsilon_i=\varepsilon_\infty=1$) we obtain, to lowest order
in $\Delta_V$, 
  \be
\alpha_{vv}=\Delta_V \, \frac{V_o}{4\pi}\,g_{v\perp} c_{0\perp}\,
\frac{\w_{0\perp}^2}{\w_{0\perp}^2-\w^2-i\w \Gamma_{\perp}}\label{eq:alpha_perp}
  \ee
in agreement with the result in
Ref. \onlinecite{Ambjornsson_Apell_Mukhopadhyay}.

The resonance condition $\Re[K_v(\kappa=1/n_v^o-1)]=0$ gives
resonance frequencies:
 \bea
(\w_{v\perp}^{\pm})^2&=&\frac{\w_p^2 n_v^o+\w_{0\perp}^2}{2}\nonumber\\
& & \hspace{-0.5cm}\pm \left[\Big(\frac{\w_p^2 n_v^o-\w_{0\perp}^2}{2}\Big) ^2
+\w_{0\perp}^4 D_{v\perp}^2 \right]^{1/2}\label{eq:Rabi_perp}   
  \eea
with the dimensionless coupling strength $D_{v\perp}$ defined by
  \be
D_{v\perp}^2 =\Delta_V g_{v\perp} c_{0\perp}n_v^o
\left(\frac{\w_p^2 n_v^o +\w_{0\perp}^2}{2\w_{0\perp}^2}\right),\label{eq:D_perp_app}
  \ee
compare to Eq. (\ref{eq:D_paral_app}).

The scaled peak height $T^+_{v\perp}$ ($T^-_{v\perp}$) of the
resonance characterized by resonance frequency $\w_{v\perp}^+$
($\w_{v\perp}^-$) are obtained from Eq. (\ref{eq:T_v_def}). We find
(for $\w_p\neq 0$)
  \bea
T^\pm_{v\perp}&=&\frac{\w_p^2}{(\w_{v\perp}^{\pm})^2}\nonumber\\
& &\hspace{-1.5cm}\times \frac{\big(\w_{0\perp}^2-(\w_{v\perp}^{\pm})^2\big)}
{\Gamma_p\big(\w_{0\perp}^2-(\w_{v\perp}^{\pm})^2\big)+\Gamma_{\perp}
\big(\w_p^2n_v^o-(\w_{v\perp}^{\pm})^2\big)}\label{eq:T_v_perp}
  \eea
in the same manner as in the preceding subsection.

We notice now from Eq. (\ref{eq:Rabi_perp}) that if $D_{v\perp}\ll
|[\w_p^2 n_v^o-\w_{0\perp}^2]/[2 \w_{0\perp}^2]|$, then the resonance
frequencies are the uncoupled ones: $\w_p\sqrt{n_v^o}$ and
$\w_{0\perp}$.

If instead $D_{v\perp}\gg |[\w_p^2 n_v^o-\w_{0\perp}^2]/[2
  \w_{0\perp}^2]|$, (i.e., for $\w_p^2 n_v^o \approx \w_{0\perp}^2$),
we get $(\w_{v\perp}^{\pm})^2 \approx \w_p^2 n_v^o[1\pm D_{v\perp}]$,
and
 \be
 T^\pm_{v\perp}= \frac{1}{n_v^o (\Gamma_p+\Gamma_{\perp})}\,\frac{1}{1\pm D_{v\perp}}
  \ee
with $D_{v\perp}=(\Delta_V g_{v\perp}c_{0\perp}n_v^o)^{1/2}$, 
showing again strong hybridization of the plasmon and coating resonances.

\subsection{Damping constants}

In the preceding subsections we derived expression for the resonance
frequencies and peak heights for the coating-nanoparticle hybridized
resonances. We here proceed by also deriving expressions for the
damping constants (inverse life times), which determine the
peak {\em widths}. It is convenient to define
  \be
\Delta \w^2 \equiv (\w_{v\mu}^{\pm})^2 -\w^2,
  \ee
where $\w_{v\mu}^{\pm}$ are the resonance frequencies, see
Eqs. (\ref{eq:Rabi_paral}) and (\ref{eq:Rabi_perp}). We can then
expand $K_v(\kappa)$ in Eq. (\ref{eq:alpha_no_coat}) according to:
  \be
K_v(\kappa)=K_v^{(0)}(\kappa)+\Delta \w^2 K_v^{(1)}(\kappa)+..,
  \ee
where $K_v^{(0)}(\kappa)=K_v(\kappa)|_{\w=\w_{v\mu}^{\pm}}$. Inserting this
expansion into the expression for the polarizability
Eq. (\ref{eq:alpha_no_coat}), using the resonance condition
$\Re[K_v(\kappa=1/n_v^o-1)]=0$, and keeping only terms to lowest order
in $\Delta \w^2$ we find
  \bea
\alpha_{vv}(\w\sim \w_{v\mu}^{\pm})\approx \frac{V_o}{4\pi n_v^o} \nonumber\\
&& \hspace{-4.5cm}\times \frac{\Re[K_v^{(0)}(-1)]+i\Im [K_v^{(0)}(-1)]}{\Delta \w^2 \Re[K_v^{(1)}(1/n_v^o-1)]+i\Im [K_v^{(0)}(1/n_v^o-1)]}.\label{eq:alpha_close_to_res}
  \eea
By furthermore neglecting the imaginary part in the numerator (i.e.,
assuming small damping) we find
  \bea
\alpha_{vv}(\w\sim \w_{v\mu}^{\pm})&\approx& \frac{V_o}{4\pi n_v^o} \frac{\Re[K_v^{(0)}(-1)]}{(\w_{v\mu}^{\pm})^2 \Re[K_v^{(1)}(1/n_v^o-1)]}\nonumber\\
& &\times \frac{(\w_{v\mu}^{\pm})^2}{(\w_{v\mu}^{\pm})^2 -\w^2 -i\w \Gamma_{v\mu}^{\pm}}\label{eq:alpha_close_to_res2},
  \eea
where we introduced the damping constant
  \be
\Gamma_{v\mu}^{\pm}=\frac{\Re[K_v^{(0)}(1/n_v^o-1)]}{\w_{v\mu}^{\pm} \Re[K_v^{(1)}(1/n_v^o-1)]}\label{eq:Gamma}.
  \ee
In order to obtain an explicit expression for the damping we proceed
by expanding Eqs. (\ref{eq:K_paral}) and (\ref{eq:K_perp}) and obtain
  \be
\Re[K_v^{(1)}(1/n_v^o-1)]=\frac{2}{n_v^o} \Big(\w_{a\mu}^2-(\w_{v\mu}^{\pm})^2 \Big)\label{eq:expansion},
  \ee
where $\w_{a\mu}^2= (\w_{0\mu}^2 +\w_p^2n_v^o)/2$, and we neglected
terms of order $\Delta_V \Delta\w^2$. Combining Eqs. (\ref{eq:Gamma})
and (\ref{eq:expansion}) we finally find that the damping constant takes the
form
  \be
\Gamma_{v\mu}^{\pm}=  \frac{\Gamma_p\big(\w_{0\mu}^2-(\w_{v\mu}^{\pm})^2\big)+\Gamma_{\mu}\big(\w_p^2n_v^o-(\w_{v\mu}^{\pm})^2\big)}{2\Big(\w_{a\mu}^2 -(\w_{v\mu}^{\pm})^2\Big)}\label{eq:Gamma_final}.
  \ee
Concluding, we above showed that close to one of the resonances the
polarizability can be approximated by a Lorentzian, see
Eqs. (\ref{eq:alpha_close_to_res2}), with a width determined by
Eq. (\ref{eq:Gamma_final}).

Using Eq. (\ref{eq:alpha_close_to_res}) and the definition
Eq. (\ref{eq:T_v_def}) we find the following relation between the
scaled peak height and the damping constant
  \be
T_{v\mu}^{\pm}=\frac{\w_p^2}{2(\w_{v\mu}^{\pm})^2} \frac{\w_{0\mu}^2-(\w_{v\mu}^{\pm})^2}{\w_{a\mu}^2-(\w_{v\mu}^{\pm})^2}\frac{1}{\Gamma_{v\mu}^{\pm}},
  \ee
to lowest order in $\Delta_V$, compare
Eq. (\ref{eq:T_v_plasmon}). Inserting Eq. (\ref{eq:Gamma_final}) into
the equation above we recover the results in Eqs. (\ref{eq:T_v_paral})
and (\ref{eq:T_v_perp}) as it should.

%%%%%%%%%%%%%%% Bibliography %%%%%%%%%%%%%%%%%%%%


\begin{thebibliography}{10}
\bibitem{cylindrical_case}
For the case of a coated cylinder ($s=1$ and $a_x\rightarrow\infty$)
one could use the results for the polarizability given in Refs.
\onlinecite{Ambjornsson_Mukhopadhyay} or \onlinecite{Henrard_Lambin}
in order to derive a corresponding expression.
\end{thebibliography}
\end{document}